\pdfoutput=1
\documentclass[11pt,draftcls,onecolumn]{IEEEtran}\newcommand{\matrixfontsize}{\fontsize{11}{11}\selectfont}

\usepackage[T1]{fontenc}
\usepackage{amsmath} 
\usepackage{amssymb} 
\usepackage[mathscr]{eucal} 
\usepackage[final]{graphicx} 
\usepackage{hyperref}
\IEEEeqnarraydefcolsep{15}{1.5\arraycolsep} 

\DeclareMathOperator{\Aut}{Aut}
\newcommand{\bsy}[1]{\boldsymbol{#1}} 
\newcommand{\diag}{\mbox{\rm diag}}
\newcommand{\gam}[1]{\mbox{$\gamma_{\!\ssst#1}$}} 
\newcommand{\order}{\mbox{\rm order}}
\newcommand{\ssst}{\scriptscriptstyle}
\newcommand{\suppint}{\mbox{\rm supp\_int}}

\newtheorem{defn}{Definition} 
\newtheorem{exmp}{Example} 
\newtheorem{lem}{Lemma} 
\newtheorem{prop}{Proposition} 
\newtheorem{thm}{Theorem}

\newcommand{\rem}{\paragraph*{Remarks}} 
\newcommand{\pf}{\IEEEproof} 
\newcommand{\qed}{\IEEEQED}  

\begin{document}
\date{December 10, 2009}
\title{Group Lifting Structures for Multirate Filter Banks,~I: Uniqueness of Lifting Factorizations}
\author{Christopher M.\ Brislawn
	\thanks{The author is with Los Alamos National Laboratory, Los Alamos, NM 87545--1663 USA (e-mail: brislawn@lanl.gov).   Los Alamos National Laboratory is operated by Los Alamos National Security LLC for the U.\ S.\ Department of Energy under contract DE-AC52-06NA25396.  This work was supported by the Los Alamos Laboratory-Directed Research \& Development Program.}\\
Los Alamos National Laboratory, MS B265, Los Alamos, NM 87545--1663 USA\\
(505) 665--1165 (office);\quad  (505) 665--5220 (FAX);\quad e-mail: {\tt brislawn@lanl.gov}
}

\maketitle
\vspace*{-0.25in}
\begin{center}
\textbf{\Large Final Revision, Reformatted for arXiv.org}
\end{center}
\vspace*{0.25in}

\begin{abstract}
Group lifting structures are introduced to provide an algebraic framework for studying lifting factorizations of two-channel perfect reconstruction FIR filter banks.  The lifting factorizations generated by a group lifting structure are characterized by abelian groups of lower and upper triangular lifting matrices, an abelian group of unimodular gain scaling matrices, and a set of base filter banks.  Examples of group lifting structures are given for  linear phase lifting factorizations of the two nontrivial classes of two-channel linear phase FIR filter banks, the whole- and half-sample symmetric classes, including both the reversible and irreversible cases.   This covers the lifting specifications for whole-sample symmetric filter banks in Parts~1 and~2 of the ISO/IEC JPEG~2000 still image coding standard.    The theory is used to address the  uniqueness of lifting factorizations.  With no constraints on the lifting process, it is shown that lifting factorizations are highly nonunique.
When certain hypotheses developed in the paper  are satisfied, however,  lifting factorizations generated by  a group lifting structure  are shown to be unique.  
A companion paper applies the uniqueness results proven in this paper to the linear phase group lifting structures for whole- and half-sample symmetric filter banks.
\end{abstract}


\newpage\tableofcontents\listoffigures\clearpage


\section{Introduction}\label{sec:Intro}
\begin{figure}[tb]
  \begin{center}
    \includegraphics{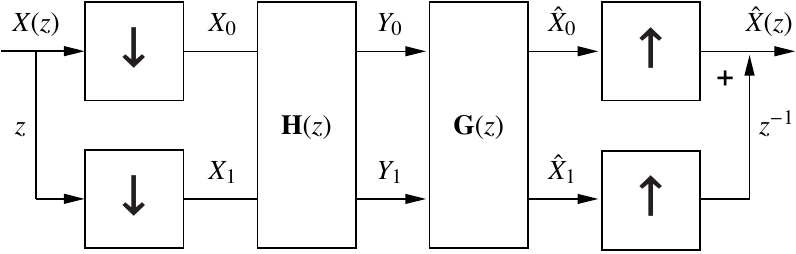}
    \caption{The polyphase-with-advance filter bank representation.}
    \label{DS_poly}
  \end{center}
\end{figure}

This paper studies two-channel finite impulse response (FIR) perfect reconstruction filter banks~\cite{Vaid93}.  Figure~\ref{DS_poly} depicts the polyphase-with-advance representation of a  filter bank~\cite{BrisWohl06}.  
A lifting factorization~\cite{Sweldens96,Sweldens98}  decomposes the polyphase matrices $\mathbf{H}(z)$ and $\mathbf{G}(z)$ into  upper and lower triangular \emph{lifting matrices}, a variant of a well-known result from linear algebra (e.g.,~\cite[Proposition~VII.2.11]{Hungerford74}) stating that any matrix over a Euclidean domain can be diagonalized by elementary matrices.  The construction of such factorizations via the Euclidean algorithm was given for general FIR perfect reconstruction filter banks in~\cite{DaubSwel98}   and was subsequently refined for  linear phase filter banks in~\cite{TaubMarc02,BrisWohl06}.  These latter works were motivated by the ISO/IEC JPEG~2000  standard~\cite{ISO_15444_1,ISO_15444_2,TaubMarc02}, which specifies whole-sample symmetric (WS, or FIR type~1 linear phase) wavelet filter banks in terms of \emph{half-}sample symmetric (HS, or FIR type~2) lifting filters.  
(The connection between filter banks and wavelet transforms is well-known and will not be treated here; see~\cite{Daub92,VettKov95,StrNgu96,Mallat99}.)  %
\begin{figure}[tb]
  \begin{center}
    \includegraphics{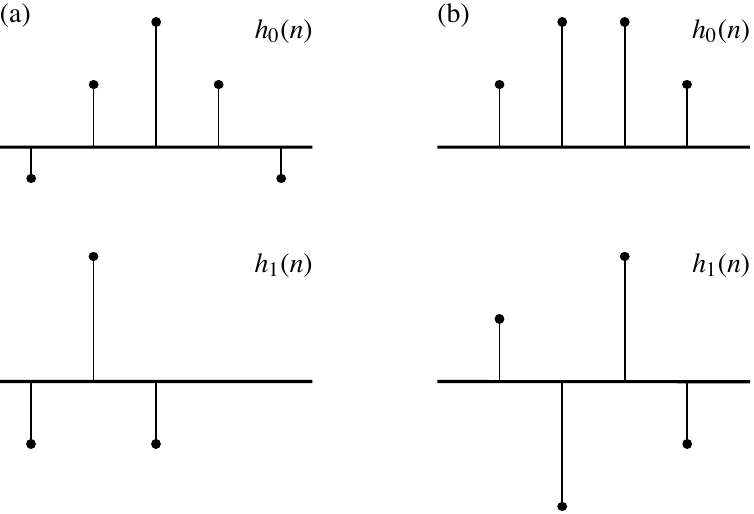}
    \caption{(a)~Whole-sample symmetric (WS) filter bank.\quad 
    (b)~Half-sample symmetric (HS) filter bank.}
    \label{WS-HS}
  \end{center}
\end{figure}

Elementary matrix decompositions are known to be nonunique and it is also known that, in general, lifting factorizations of filter banks are similarly nonunique.  Nonetheless, this paper introduces a framework for lifting factorizations, called a \emph{group lifting structure}, that allows us to prove  \emph{uniqueness} of  lifting steps under carefully stated hypotheses.  This is an extension of the group-theoretic approach to lifting introduced in~\cite{BrisWohl06}.  A companion paper~\cite{Bris09b} applies the theory developed here to the two principal classes of linear phase filter banks, the WS and HS classes shown in Figure~\ref{WS-HS}.  It is proven in~\cite{Bris09b} that there is \emph{only one way} to factor a given WS filter bank into HS lifting filters as specified in~\cite{ISO_15444_1,ISO_15444_2},  even though the given filter bank has many other lifting factorizations into different (e.g., nonlinear phase) lifting filters.  As shown in~\cite{BrisWohl06}, factoring HS filter banks into linear phase  lifting steps is more complicated than in the WS case, but~\cite{Bris09b} contains analogous results on the uniqueness of linear phase lifting factorizations for HS filter banks.  
A follow-on paper currently in preparation 
studies the group-theoretic structure of lifting factorizations  and determines the groups (up to isomorphism)  arising in unique lifting factorizations.
This  work  is being applied to develop a group-theoretic numerical framework for optimal filter bank design, a goal motivated by the observation that perfect reconstruction filter banks do not form vector spaces.

\subsection{Nonuniqueness of Lifting Factorizations}\label{sec:Intro:Nonuniqueness}
Examples of nonunique lifting factorizations are ubiquitous.  Consider our old friend, the Haar filter bank, normalized with a polyphase determinant of~1 and a lowpass DC response of~1:
\begin{equation}\label{Haar}
\mathbf{H}_{haar}(z) \equiv 
        \left[ \begin{IEEEeqnarraybox*}[\matrixfontsize][c]{,c15c,}
                1/2 & 1/2 \\
                -1 & 1
        \end{IEEEeqnarraybox*}\right]\,.
\end{equation}
The lifting factorization of~(\ref{Haar}) in~\cite{DaubSwel98} and~\cite[Annex~H]{ISO_15444_2}  is 
\begin{equation}\label{Haar_lifting_steps}
\mathbf{H}_{haar}(z) = 
        \left[ \begin{IEEEeqnarraybox*}[\matrixfontsize][c]{,c15c,}
                1  & 1/2 \\
                0  & 1
        \end{IEEEeqnarraybox*}\right]
        \left[ \begin{IEEEeqnarraybox*}[\matrixfontsize][c]{,c15c,}
                1  & 0 \\
                -1  & 1
        \end{IEEEeqnarraybox*}\right]\,.
\end{equation}
The Haar matrix has another factorization, however, with different lifting steps and a  diagonal gain scaling matrix:
\begin{equation}\label{alt_Haar_lifting_steps}
\mathbf{H}_{haar}(z) = 
        \left[ \begin{IEEEeqnarraybox*}[\matrixfontsize][c]{,c15c,}
                1/2  & 0 \\
                0    & 2
        \end{IEEEeqnarraybox*}\right]
        \left[ \begin{IEEEeqnarraybox*}[\matrixfontsize][c]{,c15c,}
                1  & 0 \\
                -1/2  & 1
        \end{IEEEeqnarraybox*}\right]
        \left[ \begin{IEEEeqnarraybox*}[\matrixfontsize][c]{,c15c,}
                1  & 1 \\
                0  & 1
        \end{IEEEeqnarraybox*}\right]\,.
\end{equation}
Similar observations are made in~\cite{DaubSwel98} regarding nonunique factorizations of the 4-tap/4-tap (${D}_4$) orthogonal wavelet filter bank, and that example can be extended to exhibit nonunique lifting factorizations for the entire family of orthogonal Daubechies (${D}_{2n}$) wavelet filter banks.

To appreciate just how fundamentally nonunique lifting factorizations are, equate~(\ref{Haar_lifting_steps}) and~(\ref{alt_Haar_lifting_steps}) and move the lifting steps from~(\ref{Haar_lifting_steps}) over to the right end of~(\ref{alt_Haar_lifting_steps}). Use~\cite[Section~7.3]{DaubSwel98} to factor $\diag(1/2,\,2)$  into lifting steps, yielding factorization~(\ref{identity_lift}).
\begin{IEEEeqnarray}{rCl}
\mathbf{I}&=&
        \left[ \begin{IEEEeqnarraybox*}[\matrixfontsize][c]{,c15c,}
                1 & 0 \\
               -1 & 1
        \end{IEEEeqnarraybox*}\right]
        \left[ \begin{IEEEeqnarraybox*}[\matrixfontsize][c]{,c15c,}
                1 & -1 \\
                0 & 1
        \end{IEEEeqnarraybox*}\right]
        \left[ \begin{IEEEeqnarraybox*}[\matrixfontsize][c]{,c15c,}
                1       & 0 \\
                1/2    & 1
        \end{IEEEeqnarraybox*}\right]
        \left[ \begin{IEEEeqnarraybox*}[\matrixfontsize][c]{,c15c,}
                1   & 2 \\
                0   & 1
        \end{IEEEeqnarraybox*}\right] \IEEEnonumber\\%
&&\cdot\left[ \begin{IEEEeqnarraybox*}[\matrixfontsize][c]{,c15c,}
                1  & 0 \\
                -1/2  & 1
        \end{IEEEeqnarraybox*}\right]
        \left[ \begin{IEEEeqnarraybox*}[\matrixfontsize][c]{,c15c,}
                1  & 1 \\
                0  & 1
        \end{IEEEeqnarraybox*}\right]
       \left[ \begin{IEEEeqnarraybox*}[\matrixfontsize][c]{,c15c,}
                1  & 0 \\
                1  & 1
        \end{IEEEeqnarraybox*}\right]
        \left[ \begin{IEEEeqnarraybox*}[\matrixfontsize][c]{,c15c,}
                1  & -1/2 \\
                0  & 1
        \end{IEEEeqnarraybox*}\right]
        \label{identity_lift}
\end{IEEEeqnarray}
The formulas  in~\cite[Section~7.3]{DaubSwel98} can be used to  generate infinitely many other lifting factorizations of the identity.

In spite of the general nonuniqueness of elementary matrix decompositions,  work on the JPEG~2000 standard  raised a number of questions regarding uniqueness of \emph{linear phase} lifting factorizations.  For instance, the lifting factorization of WS  filter banks in~\cite[Theorem~9]{BrisWohl06} involves left matrix ``downlifting'' of a WS transfer matrix into  HS lifting steps.  One is naturally led to ask whether the same sequence of HS lifting filters is obtained for right matrix factorizations or, more generally,  for a mix of  left and right factorization operations.   

For example, the software accompanying~\cite{Maslen97,MaslenAbbott:00:Automation-lifting-factorisation} uses the Euclidean algorithm to  compute nine different lifting factorizations of a 7-tap/5-tap Cohen-Daubechies-Feauveau WS wavelet filter bank based on cubic B-splines~\cite{CohenDaubFeau92,Daub92}.  One factorization~(\ref{CDF7-5}), computed in~\cite[Section~5.4]{MaslenAbbott:00:Automation-lifting-factorisation} in synthesis matrix form,
uses HS lifting filters exclusively (the other eight factorizations involve non-symmetric lifting filters).
\begin{IEEEeqnarray}{l}
        2^{-9/2}\left[ \begin{IEEEeqnarraybox*}[\matrixfontsize][c]{,c15c,}
                -12z+40-12z^{-1}  & -8-8z^{-1} \IEEEstrut[10pt][0pt]\\
                3z^2+5z+5+3z^{-1}  & 2z+12+2z^{-1}\IEEEstrut[12pt][0pt]
        \end{IEEEeqnarraybox*}\right] \IEEEnonumber\\%
        \qquad \qquad \qquad =
        \left[ \begin{IEEEeqnarraybox*}[\matrixfontsize][c]{,c15c,}
                1  & 0\IEEEstrut[10pt][0pt] \\
                -(z+1)/4  & 1\IEEEstrut[12pt][0pt]
        \end{IEEEeqnarraybox*}\right]
        \left[ \begin{IEEEeqnarraybox*}[\matrixfontsize][c]{,c15c,}
                1  & -1-z^{-1} \IEEEstrut[10pt][0pt]\\
                0  & 1\IEEEstrut[12pt][0pt]
        \end{IEEEeqnarraybox*}\right]\label{CDF7-5}\\%
        \qquad \qquad \qquad\quad\cdot\left[ \begin{IEEEeqnarraybox*}[\matrixfontsize][c]{,c15c,}
                1  & 0 \IEEEstrut[10pt][0pt]\\
                3(z+1)/16  & 1\IEEEstrut[12pt][0pt]
        \end{IEEEeqnarraybox*}\right] 
        \left[ \begin{IEEEeqnarraybox*}[\matrixfontsize][c]{,c15c,}
                2\sqrt{2}  & 0 \\
                0    & 1/2\sqrt{2}
        \end{IEEEeqnarraybox*}\right] \IEEEnonumber
\end{IEEEeqnarray}
This exact same linear phase lifting is obtained using both left and right matrix downlifting factorization as in~\cite{BrisWohl06}.

Other questions about unique factorization arise when factoring HS filter banks, which contain an HS lowpass filter and a half-sample \emph{anti}-symmetric (HA, or FIR type~4) highpass filter, as in Figure~\ref{WS-HS}(b).  HS filter banks factor using whole-sample antisymmetric (WA, or FIR type~3) lifting filters and concentric equal-length HS base filter banks~\cite[Section~VI]{BrisWohl06}.
For instance,  a 6-tap/10-tap HS wavelet filter bank, $\mathbf{H}_{6,10}(z)$, can be derived from the same halfband spectral factorization as the 9-tap/7-tap Cohen-Daubechies-Feauveau WS wavelet filter bank in~\cite{CohenDaubFeau92},~\cite[Figure~8.8]{Daub92}.  
Annex~H.4.1.2.1 of~\cite{ISO_15444_2} has a  factorization of $\mathbf{H}_{6,10}(z)$  in which it is lifted from a concentric 6-tap/6-tap HS base filter bank by a second-order WA lifting step.    
At the time that~\cite{ISO_15444_2} was written, it was unknown whether it was possible to lift $\mathbf{H}_{6,10}(z)$ from the Haar filter bank using WA lifting filters.
The same question was raised about a 10-tap/18-tap HS filter bank that is lifted from a 10-tap/10-tap HS base filter bank in~\cite[Annex~H.4.1.2.2]{ISO_15444_2} by a fourth-order WA lifting filter. 
It would have simplified the writing of~\cite[Annex~H]{ISO_15444_2}  if all HS filter banks could be lifted by WA lifting filters from a single base filter bank such as the Haar, but our results will show this to be impossible.

\subsection{Technical Approach}\label{sec:Intro:Approach}
Our approach, which is algebraic rather than algorithmic, ignores the mechanism used to \emph{generate} lifting factorizations (e.g., the Euclidean algorithm in~\cite{DaubSwel98} or the ``downlifting'' approach in~\cite{BrisWohl06}) and instead constrains the universe of allowable lifting steps. Under suitable hypotheses, we shall  prove uniqueness within precisely defined classes of lifting factorizations.  For example, using the results of this paper, we prove in~\cite{Bris09b} that one always obtains the same HS lifting filters when factoring a WS filter bank into HS lifting steps, regardless of how the factorization is constructed.   In~\cite{Bris09b} we also prove uniqueness of the WA lifting filters and the equal-length HS base filter bank that arise in the factorization of a suitably normalized HS filter bank.  Both results follow from Theorem~\ref{thm:unique_factorization} of the present paper, a result that  does not depend on linear phase characteristics.   
As in~\cite{BrisWohl06}, we make extensive use of the group-theoretic structure of lifting factorizations.   

Readers may be surprised by the absence of traditional ring-theoretic methods for establishing unique factorization results over polynomial domains.  While ring-theoretic techniques such as Gr\"obner bases~\cite{LebrunSelesnic:04:Grobner-bases-wavelet}
have been used for {constructing} wavelet filter banks, filter banks are elements of \emph{matrix} polynomial rings, which are not integral domains.  More significantly, {perfect reconstruction} filter banks (including lifting matrices) are \emph{units} in the ring of  $2\times 2$ matrices over the Laurent polynomials, $\mathbb{C}[z,\,z^{-1}]$.  This means that traditional algebraic methods based on ideal theory can tell us nothing about uniqueness of  lifting factorizations. 
Of course, it is also surprising  that unique factorization results can be proven using group-theoretic techniques since everything divides everything in a group, but this only highlights the importance of carefully formulating the universe of admissible lifting factors for a given class of polyphase matrices.

The author considered recasting the problem in terms of rings of \emph{causal} matrix polynomials, but doing so would have required reworking much of the basic theory of lifting and would have distracted attention from the results presented here.  Fortunately, the group-theoretic approach proved adequate for deriving the results of interest, and it probably resulted in a more elementary level of algebraic machinery than would have been used in a ring-theoretic approach.  Moreover, the technical crux of the group-theoretic approach, which is the proof of an order-increasing property, would probably be required in a ring-theoretic treatment as well.
The reader should note that this paper does not require an extensive background in group theory.  The basic material---group axioms, subgroups, and homomorphisms---found in any introductory text on ``abstract algebra'' will suffice; e.g.,~\cite{Hungerford74,MacLaneBirkhoff67,Jacobson74,Herstein75}.

\subsubsection{Comparison to Other Work}\label{sec:Intro:Compare}
Daubechies and Sweldens~\cite{DaubSwel98} point out that the Euclidean algorithm can generate many different lifting factorizations for a given filter bank, and they pose the analysis and exploitation of nonuniqueness as ``an interesting topic for future research.''  Nonuniqueness of lifting factorizations  was studied in~\cite{StefanoiTabus:02:Euclidean-lifting-schemes}
to find optimal integer-to-integer liftings for a given reversible filter bank.  Recent research by Zhu and Wickerhauser~\cite{ZhuWickerhauser_FFT08} exploits nonuniqueness  to construct liftings that use only ``nearest-neighbor'' data (i.e., first-order lifting filters of the form $a+bz^{-1}$ or $az+b$) in order to minimize memory fetches in hardware implementations.  Their work shows that some lifting factorizations are much more ill-conditioned than others, which likely affects rate-distortion scalability in source coding applications.  As far as the author is aware, however, there are no prior results establishing \emph{uniqueness} of lifting steps for \emph{any} classes of lifting factorizations.

Unlike~\cite{FooteMirchandEtal:00:Wreath-Product-Group,MirchandFooteEtal:00:Wreath-Product-Group,FooteMirchandEtal:04:Two-Dimensional-Wreath-Product}, 
which also apply  group theory to  multirate discrete-time systems, the present paper makes no use of the specialized theory of \emph{finite} groups.  Moreover, our results are not constrained to  finite signals of length $2^n$: the papers~\cite{FooteMirchandEtal:00:Wreath-Product-Group,MirchandFooteEtal:00:Wreath-Product-Group,FooteMirchandEtal:04:Two-Dimensional-Wreath-Product} 
use group-theoretic tools to analyze the multiscale structure of the underlying space of finite-length signals, whereas we employ group-theoretic methods to analyze the algebraic structure of multirate FIR filter banks acting on  general discrete-time signals with no length constraints.

\subsubsection{Outline Of The Paper}\label{sec:Intro:Outline}
Section~\ref{sec:Filter} reviews filter banks and lifting.  Section~\ref{sec:Abelian} introduces the group-theoretic concepts that provide the framework for our uniqueness results, which are based on the notion of \emph{group lifting structures}.  Section~\ref{sec:LinearPhase} presents examples of group lifting structures for factoring WS and HS filter banks into linear phase lifting filters, including the ELASF class of reversible HS filter banks lifted from the Haar filter bank that was defined by M.\ Adams.  In Section~\ref{sec:Uniqueness} we define the key \emph{polyphase order-increasing property} that a group lifting structure must satisfy to ensure unique lifting factorizations, and we prove the main uniqueness theorem.    Section~\ref{sec:Concl} contains concluding remarks.

\section{Filter Banks and Lifting Factorizations}\label{sec:Filter}
This section reviews notation and introduces several important concepts and tools for working with lifting factorizations.  

\subsection{Filters, Filter Banks, and Support Intervals}\label{sec:Filter:SupportIntervals}
Throughout the paper,  FIR filters are written as (possibly noncausal) Laurent polynomials,
\[ F(z) \equiv \sum_{n=a}^b f(n)\,z^{-n}\in \mathbb{C}\left[z,z^{-1}\right]\,. \]
\begin{defn}\label{defn:SupportInterval}
The {\em support interval\/} of an FIR filter, denoted
\begin{equation}\label{supp_int}
\suppint(F)\equiv \suppint(f)\equiv [a,b]\subset\mathbb{Z}\;, 
\end{equation}
is the smallest closed interval of integers containing the support of the filter's impulse response or, equivalently, the largest closed interval for which $f(a)\neq 0$ and \mbox{$f(b)\neq 0$.}
If 
\[ \suppint(f)=[a,b] \]
then the {\em order\/} of $F(z)$ is  
\begin{equation}\label{order}
\order(F) \equiv b-a\,.
\end{equation}
\end{defn}

The analysis polyphase-with-advance decomposition of a filter~\cite[eqn.~(9)]{BrisWohl06} is 
\[ F(z)=F_0(z^2) + zF_1(z^2) \,, \]
and its \emph{analysis polyphase vector representation}  is
\begin{IEEEeqnarray}{rCl}
\bsy{F}(z) & \equiv &
        \left[ \begin{IEEEeqnarraybox*}[\matrixfontsize][c]{,c,}
        F_0(z)\IEEEstrut[10pt][3pt]\\
        F_1(z)\IEEEstrut[10pt][3pt]
        \end{IEEEeqnarraybox*}\right]
=  \sum_{n=c}^{d}\bsy{f}(n)\,z^{-n},\label{poly_fltr}\\
\bsy{f}(n) & \equiv & 
        \left[ \begin{IEEEeqnarraybox*}[\matrixfontsize][c]{,c,}
        f_0(n)\IEEEstrut[10pt][3pt]\\
        f_1(n)\IEEEstrut[10pt][3pt]
        \end{IEEEeqnarraybox*}\right]
\quad\mbox{with $\bsy{f}(c),\,\bsy{f}(d)\neq\bsy{0}$\,.}\label{poly_fltr_impulse}
\end{IEEEeqnarray}
In this paper we generally work with analysis (rather than synthesis) polyphase-with-advance filter bank representations, so henceforth ``polyphase'' will mean ``analysis polyphase-with-advance'' unless stated otherwise.
The polyphase  representation~(\ref{poly_fltr}), (\ref{poly_fltr_impulse}) has the \emph{polyphase support interval}
\begin{equation}\label{poly_supp_int}
\suppint(\bsy{f}) \equiv [c,d] \,, 
\end{equation}
which differs from the scalar support interval~(\ref{supp_int}) for  the same filter.
The \emph{polyphase order} of~(\ref{poly_fltr}) is
\begin{equation}\label{poly_filt_order}
\mbox{order}(\bsy{F}) \equiv d-c\;.
\end{equation}

These definitions generalize for FIR filter banks, $\{H_0(z),\,H_1(z)\}$.  Decompose each filter, $H_i(z)$, $i=0,\,1$, into its polyphase vector representation, $\bsy{H}_i(z)$, as in~(\ref{poly_fltr}) and form the polyphase matrix:
\begin{IEEEeqnarray}{rCl}
\mathbf{H}(z) & \equiv  & 
        \left[ \begin{IEEEeqnarraybox*}[\matrixfontsize][c]{,c,}
        \bsy{H}_0^T(z)\IEEEstrut[10pt][3pt]\\
        \bsy{H}_1^T(z)\IEEEstrut[11pt][3pt]
        \end{IEEEeqnarraybox*}\right]
= \sum_{n=c}^d \mathbf{h}(n)\,z^{-n}\,,\label{poly_matrix_def}\\
\mathbf{h}(n) & \equiv & 
        \left[ \begin{IEEEeqnarraybox*}[\matrixfontsize][c]{,c,}
        \bsy{h}_0^T(n)\IEEEstrut[10pt][3pt]\\
        \bsy{h}_1^T(n)\IEEEstrut[11pt][3pt]
        \end{IEEEeqnarraybox*}\right]
\quad\mbox{with $\mathbf{h}(c),\,\mathbf{h}(d)\neq \mathbf{0}$\,.}\label{poly_matrix_impulse_def}
\end{IEEEeqnarray}
Note that we use bold italics for column vectors and bold roman (upright) fonts for matrices.
The polyphase support interval of the filter bank in~(\ref{poly_matrix_def}), (\ref{poly_matrix_impulse_def}) is defined to be
\begin{equation}
\suppint(\mathbf{h}) \equiv [c,d] \label{FB_suppint}
\end{equation}
and the polyphase order  is
\begin{equation}
\mbox{order}(\mathbf{H}) \equiv d-c\,.\label{FB_order}
\end{equation}

\subsection{Lifting}\label{sec:Filter:Lifting}
$\mathbf{H}(z)$ is an \emph{FIR perfect reconstruction filter bank} (i.e., an FIR filter bank with FIR inverse) if and only if 
\[ \det\mathbf{H}(z) = az^{-d}\quad\mbox{for $a\neq 0$, $d\in\mathbb{Z}$.} \]
The set of all such filter banks forms a multiplicative group, $\mathscr{F}$,  called the \emph{FIR filter bank group}~\cite[Theorem~1]{BrisWohl06}.  A polyphase matrix is called \emph{unimodular}~\cite{MacLaneBirkhoff67} if
\begin{equation}\label{DS_FIR_PR}
\det\mathbf{H}(z) = 1\,.
\end{equation}
In~\cite[Theorem~1]{BrisWohl06} it was noted that the filter banks satisfying~(\ref{DS_FIR_PR}) form a normal subgroup, $\mathscr{N\vartriangleleft F}$, of the FIR filter bank group.      We refer to $\mathscr{N}$,  which readers may recognize as \mbox{$SL(2,\, \mathbb{C}\left[z,z^{-1}\right])$}, as the \emph{unimodular group}.  (This meaning of  \emph{unimodular} is different from its use in group representation theory, where a group is called unimodular if its left and right translation-invariant measures are the same.)

\begin{figure}[tb]
  \begin{center}
    \includegraphics{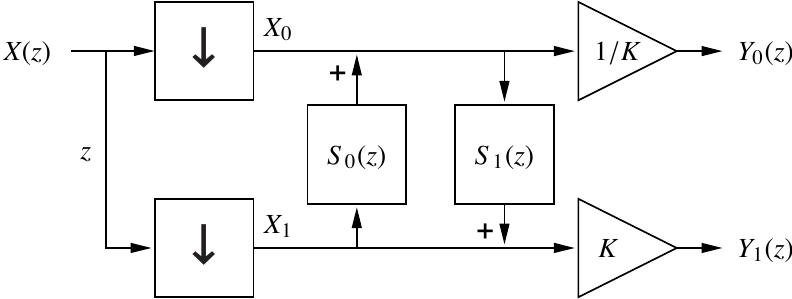}
    \caption{Lifting form of a two channel irreversible analysis filter bank.}
    \label{irr_anal_lift}
  \end{center}
\end{figure}

Daubechies and Sweldens~\cite{DaubSwel98} proved that any unimodular FIR transfer matrix has a {\em lifting factorization}:
\begin{equation}\label{anal_lift_cascade}
\mathbf{H}(z) = \diag(1/K,\,K)\,\mathbf{S}_{N-1}(z)\cdots\mathbf{S}_1(z)\,\mathbf{S}_0(z)\;.
\end{equation}
We also refer to~(\ref{anal_lift_cascade}) as a  \emph{lifting cascade}.   The diagonal matrix, $\diag(1/K,\,K)$, is a \emph{unimodular gain scaling matrix} with \emph{scaling factor} $K\neq 0$.  The lifting matrices $\mathbf{S}_i(z)$ are upper or lower triangular with ones on the diagonal and a  lifting filter, $S_i(z)$, in  the  off-diagonal position.  In the factorization corresponding to  Figure~\ref{irr_anal_lift}, for instance, the lifting matrix for the step containing ${S}_0(z)$ (which is a lowpass update) is upper triangular,  and the matrix  for the second step (a highpass update) is lower triangular.

\begin{defn}\label{defn:update_char}
The {\em update characteristic} of a lifting step (or lifting matrix) is a binary flag, $m=0$ or $1$, indicating which polyphase channel is being updated by the lifting step.  
\end{defn}

For instance, the update characteristic, $m_0$, of the first lifting step in Figure~\ref{irr_anal_lift}  is  ``lowpass,'' which is coded with a zero: $m_0=0$.  Similarly, the update characteristic of the second lifting step is ``highpass'' (coded with a one: $m_1=1$).  The update characteristic, denoted $m_i$, is defined similarly for each lifting matrix, $\mathbf{S}_i(z)$, in a lifting cascade~(\ref{anal_lift_cascade}).  (The term ``update characteristic'' is taken from~\cite[Annex~G]{ISO_15444_2}.)

\subsubsection{Partially Factored Lifting Cascades}
\label{sec:Filter:Lifting:Partial}
To  cover  cases that lift one filter bank to another,  we do not assume all lifting cascades start from the identity (i.e., are  ``lifted from the lazy wavelet'').  Instead, we often work with \emph{partially factored} lifting cascades relative to some \emph{base} filter bank  $\mathbf{B}(z)$,
\begin{equation}\label{lift_from_B}
\mathbf{H}(z) = \diag(1/K,\,K)\,\mathbf{S}_{N-1}(z) \cdots\mathbf{S}_0(z)\, \mathbf{B}(z)\,,
\end{equation}
with corresponding scalar filters $B_0(z)$ and $B_1(z)$.
Since the lifting matrices multiply $\mathbf{B}(z)$ on the left, (\ref{lift_from_B}) should be called a  {\em left\/} lifting cascade.  This is the primary case of interest, however, so ``partially factored lifting cascade'' will mean {\em left\/} cascade, as in~(\ref{lift_from_B}), unless specified otherwise.
Moreover, we usually ignore the lifting steps involved in factoring $\mathbf{B}(z)$  and only study the steps $\mathbf{S}_i(z)$ to the left of $\mathbf{B}(z)$ in~(\ref{lift_from_B}).  Cascades like~(\ref{anal_lift_cascade}) that  are completely factored into lifting steps are regarded as the special case $\mathbf{B}(z)=\mathbf{I}$.

Lifting cascades will often be written recursively:
\begin{IEEEeqnarray}{rCl}
\mathbf{H}(z) &=&  \diag(1/K,\,K)\,\mathbf{E}^{(N-1)}(z)\,,\IEEEnonumber\\
\mathbf{E}^{(n)}(z) &=& \mathbf{S}_n(z)\,\mathbf{E}^{(n-1)}(z)
		\mbox{\quad for $0\leq n< N$,}\label{recursive_cascade}\\
\mathbf{E}^{(-1)}(z)  &\equiv&  \mathbf{B}(z)\,.\IEEEnonumber
\end{IEEEeqnarray}
This recursive formulation is the basis for the definition of the key polyphase order-increasing property in Section~\ref{sec:Uniqueness:Order-Increasing}.

\subsection{Irreducible  Lifting Cascades}
\label{sec:Filter:Irreducible}
We can easily rule out a couple of degenerate cases that preclude uniqueness of lifting factors.
\begin{defn}\label{defn:Irreducible}
A  lifting cascade~(\ref{lift_from_B}) is {\em irreducible\/} if all  lifting steps are nontrivial ($\mathbf{S}_i(z)\neq \mathbf{I}$) and there are no consecutive lifting matrices with the same update characteristic (i.e., they strictly alternate between lower and upper triangular). 
\end{defn}

For example, filter banks are signaled in JPEG~2000 Part~2 codestreams as irreducible lifting cascades lifted from the identity ($\mathbf{B}(z)=\mathbf{I}$).
 
Irreducibility is an attribute of lifting \emph{cascades}  rather than an attribute of individual matrices, making it qualitatively different from the notion of irreducible \emph{elements} in a ring.
Requiring irreducibility poses no loss of generality since
any lifting cascade can be simplified into an irreducible cascade  using matrix multiplication.  While irreducibility eliminates numerous trivial obstacles to uniqueness of lifting factors, irreducibility alone does not  imply uniqueness, as the preceding examples show.
In fact, by inserting  factorizations of the identity such as~(\ref{identity_lift})  into a lifting of a given FIR perfect reconstruction filter bank, it follows that every filter bank has infinitely many different irreducible lifting factorizations.

Such extreme  nonuniqueness also  trivializes questions like: ``Can $\mathbf{H}(z)$ be lifted (irreducibly) from the Haar filter bank?''    Just create a partial lifting factorization  like, e.g.,
\begin{quote}
[any lifting of $\mathbf{H}$]\,[inverse of~(\ref{Haar_lifting_steps})]\,[lifting~(\ref{identity_lift})]\,$\mathbf{H}_{haar}$
\end{quote}
and simplify to irreducible form.  
This  trivial example can be generalized to yield the following pathological result. 
\begin{prop}\label{prop:Nonuniqueness}
If $\mathbf{G}(z)$ and $\mathbf{H}(z)$ are any FIR perfect reconstruction filter banks then $\mathbf{G}(z)$ can be irreducibly lifted from $\mathbf{H}(z)$  in infinitely many different ways.
\end{prop}

\subsection{Irrelevance of Zeroth-Order Lifting Filters}
\label{sec:Filter:Irrelevance}
After comparing the above nonuniqueness results to the  formulas in~\cite[Section~7.3]{DaubSwel98} for factoring unimodular gain scaling matrices into lifting steps,
it is tempting to dismiss the examples in Section~\ref{sec:Intro:Nonuniqueness}  and Proposition~\ref{prop:Nonuniqueness} (as well as the  nonunique factorizations of orthogonal filter banks in~\cite{DaubSwel98}, which also involve  zeroth-order lifting filters)  as a pathology somehow resulting from the use of constant or zeroth-order lifting filters, which are units in $\mathbb{C}\left[z,z^{-1}\right]$.
To see that the failure of uniqueness cannot be attributed to the use of zeroth-order  lifting filters, we have the following example.
\begin{exmp}\label{exmp:nonunique_3_factors}
For all $b,\,c\in\mathbb{C}$ with $b,\,c\neq 0$, $b^2\neq c^2$, the fourth-order unimodular transfer matrix
\[
\mathbf{H}(z) \equiv 
\left[ \begin{array}{cc}
b^{-2}z^2   & b^{-2}z^2(1-bz^{-1}) \IEEEstrut[11pt][5pt]\\
\frac{b^2-c^2}{b^2}(1+bz^{-1})   & \frac{b^2-c^2}{b^2}\left(1 - \frac{b^2c^2}{c^2-b^2}z^{-2}\right)\IEEEstrut[14pt][8pt]
\end{array}\right]
\]
has two different irreducible lifting factorizations,
\begin{equation}\label{nonunique_3_factors}
\mathbf{H}(z) =
\mathbf{S}_2(z)\,\mathbf{S}_1(z)\,\mathbf{S}_0(z)
 = 
\mathbf{S}'_2(z)\,\mathbf{S}'_1(z)\,\mathbf{S}'_0(z)\,. 
\end{equation}
Specifically, $\mathbf{S}_2(z)$, $\mathbf{S}_0(z)$, and $\mathbf{S}'_1(z)$ are lower triangular and $\mathbf{S}_1(z)$, $\mathbf{S}'_2(z)$, and $\mathbf{S}'_0(z)$ are upper triangular.  The lifting filters are all first-order Laurent polynomials:
\begin{IEEEeqnarray*}{lCl}
{S}_0(z) = 1+bz^{-1}\,, & \quad & {S}'_0(z) = \frac{c^2}{c^2-b^2}(1-bz^{-1})\,,\\ 
{S}_1(z) = b^{-2}z^2(1-bz^{-1})\,, & \quad & {S}'_1(z) = \frac{b^2-c^2}{b^2}(1+bz^{-1})\,,\\
{S}_2(z) = -c^2z^{-2}(1+bz^{-1})\,, & \quad & {S}'_2(z) = \frac{z^2}{b^2-c^2}(1-bz^{-1})\,.
\end{IEEEeqnarray*}
\end{exmp}

Taking $b=1$, $c=\sqrt{5}$ in~(\ref{nonunique_3_factors}) leads to a reversible lifting of the identity~(\ref{identity_lift2}) with linear phase lifting filters.
\begin{IEEEeqnarray}{rCl}
\mathbf{I}&=&
        \left[ \begin{IEEEeqnarraybox*}[\matrixfontsize][c]{,c15c,}
                1   & (-{z^2}/4)(1-z^{-1}) \IEEEstrut[9pt][0pt] \\
                0   & 1\IEEEstrut[9pt][0pt]
        \end{IEEEeqnarraybox*}\right]
        \left[ \begin{IEEEeqnarraybox*}[\matrixfontsize][c]{,c15c,}
                1       & 0 \IEEEstrut[9pt][0pt] \\
                -4(1+z^{-1})    & 1\IEEEstrut[9pt][0pt] 
        \end{IEEEeqnarraybox*}\right] \IEEEnonumber\\%
        && \cdot
        \left[ \begin{IEEEeqnarraybox*}[\matrixfontsize][c]{,c15c,}
                1 & (5/4)(1-z^{-1}) \IEEEstrut[9pt][0pt]  \\
                0 & 1\IEEEstrut[9pt][0pt] 
        \end{IEEEeqnarraybox*}\right]
        \left[ \begin{IEEEeqnarraybox*}[\matrixfontsize][c]{,c15c,}
                1 & 0 \IEEEstrut[9pt][0pt] \\
                -(1+z^{-1}) & 1\IEEEstrut[9pt][0pt] 
        \end{IEEEeqnarraybox*}\right] \label{identity_lift2} \\
        && \cdot
        \left[ \begin{IEEEeqnarraybox*}[\matrixfontsize][c]{,c15c,}
                1 & -z^2(1-z^{-1}) \IEEEstrut[9pt][0pt] \\
                0 & 1\IEEEstrut[9pt][0pt] 
        \end{IEEEeqnarraybox*}\right]
        \left[ \begin{IEEEeqnarraybox*}[\matrixfontsize][c]{,c15c,}
                1 & 0 \IEEEstrut[9pt][0pt] \\
                5z^{-2}(1+z^{-1}) & 1\IEEEstrut[9pt][0pt] 
        \end{IEEEeqnarraybox*}\right]  \IEEEnonumber
\end{IEEEeqnarray}
In short, without some nontrivial, carefully crafted limits on the scope of allowable lifting operations, there is no hope of obtaining unique lifting factorizations.

\section{Abelian Lifting Matrix Groups and\\Group Lifting Structures}\label{sec:Abelian}
A uniqueness theorem is always given relative to some universe of feasible solutions; i.e., it states that there is only one solution \emph{in that universe}.  This makes it necessary to preface a unique factorization theorem  by  specifying the universe of factors with respect to which factorizations are unique.
In this section,  universes of lifting factorizations will be defined in terms of abelian groups of lifting matrices,  abelian groups of gain scaling matrices, and  sets (not generally  groups) of base filter banks.

\subsection{Abelian Groups of Lifting Matrices}\label{sec:Abelian:Groups}
Define one-to-one mappings 
\[ \upsilon,\,\lambda : \mathbb{C}[z,z^{-1}]\rightarrow\mathscr{N} \]
that map the lifting filter $S(z)\in\mathbb{C}[z,z^{-1}]$ to lifting matrices:
\begin{equation}\label{monomorphisms}
\upsilon(S) \equiv
        \left[ \begin{IEEEeqnarraybox*}[\matrixfontsize][c]{,c15c,}
                1  & S(z) \\
                0  & 1\IEEEstrut[10pt][1pt]
        \end{IEEEeqnarraybox*}\right]
        \quad\mbox{and}\quad
\lambda(S) \equiv
        \left[ \begin{IEEEeqnarraybox*}[\matrixfontsize][c]{,c15c,}
                1  & 0\IEEEstrut[10pt][1pt]\\
                S(z)  & 1\IEEEstrut[10pt][1pt] 
        \end{IEEEeqnarraybox*}\right]\,.
\end{equation}
It is easy to show that $\upsilon$ and $\lambda$ are homomorphisms from the additive group $\mathbb{C}[z,z^{-1}]$ into the multiplicative group $\mathscr{N}$.

\begin{lem}\label{lem:GroupIso}
For any additive subgroup of the Laurent  polynomials, $\mathscr{P}<\mathbb{C}[z,z^{-1}]$, the images
$\upsilon(\mathscr{P})$ and $\lambda(\mathscr{P})$
are abelian subgroups of $\mathscr{N}$ that are isomorphic to $\mathscr{P}$.
\end{lem}
\begin{defn}\label{defn:lifting_matrix_groups}
For additive groups of Laurent polynomials $\mathscr{P}_i<\mathbb{C}[z,z^{-1}]$, \mbox{$i=0,\,1$,}
the groups $\mathscr{U}\equiv\upsilon(\mathscr{P}_0)$ and $\mathscr{L}\equiv\lambda(\mathscr{P}_1)$ are called the \emph{lifting matrix groups} generated by $\mathscr{P}_0$ and $\mathscr{P}_1$. 
Note that $\mathscr{U}$ and $\mathscr{L}$  need not be isomorphic to one another; i.e., it may be that $\mathscr{P}_0\not\cong\mathscr{P}_1$.
\end{defn}

\subsection{Gain Scaling Matrices and Inner Automorphisms}\label{sec:Abelian:Gain}
Another component of lifting is an abelian \emph{gain scaling group}, $\mathscr{D}$, of unimodular constant matrices,
\begin{equation}\label{gain_matrix}
\mathbf{D}_K \equiv \diag(1/K,\,K)\quad\mbox{for scaling factor $K\neq 0$,}
\end{equation}
with the product $\mathbf{D}_K \mathbf{D}_J = \mathbf{D}_{KJ}$.  
$\mathscr{D}$ need not contain the matrices~(\ref{gain_matrix}) corresponding to \emph{all} $K\neq 0$; $\mathscr{D}$ could be defined in terms of any multiplicative subgroup of 
\[ \mathbb{R}^*\equiv \mathbb{R}\backslash\{0\} \quad\mbox{or}\quad 
\mathbb{C}^*\equiv \mathbb{C}\backslash\{0\}\,. \]

Let $\mathscr{D}$ act on $\mathscr{N}$ via inner automorphisms:
for a given  $\mathbf{D}\in\mathscr{D}$ define the inner automorphism $\gam{\,\mathbf D}\in\Aut(\mathscr{N})$,
\begin{equation}\label{conjugation_operator}
\gam{\,\mathbf D}\mathbf{A}(z) \equiv \mathbf{D}\,\mathbf{A}(z)\,\mathbf{D}^{-1} .
\end{equation}
This is equivalent to the intertwining relation
\begin{equation}\label{intertwine_gain}
\mathbf{D}\,\mathbf{A}(z) = (\gam{\,\mathbf D}\mathbf{A}(z))\,\mathbf{D} \,.
\end{equation}
Homomorphisms  commute with inversion so there is no ambiguity in writing
\begin{equation}\label{gamma_Ainv} 
\gam{\,\mathbf D}\mathbf{A}^{-1}(z) \equiv \gam{\,\mathbf D}\left(\mathbf{A}^{-1}(z)\right) = 
\left(\gam{\,\mathbf D}\mathbf{A}(z)\right)^{-1}.
\end{equation}

We will write  ``$\gam{K}$'' as a shorthand for $\gam{\,\mathbf{D}_{\!K}}$.  Note that $\gam{K}$ is the \emph{inverse} of the ``gain conjugation operator,'' $C_K$, defined in~\cite[Section~IV-A.1]{BrisWohl06}.  This makes the mapping 
\[ \gamma:\mathbf{D}_K\mapsto \gam{K} \]
a homomorphism of $\mathscr{D}$ onto a subgroup $\gamma(\mathscr{D})<\Aut(\mathscr{N})$.  
Applying $\gam{K}$ to a generic matrix gives~(\ref{lifting_conjugation}):
\begin{equation}\label{lifting_conjugation}
\gam{K}  \left[ \begin{IEEEeqnarraybox*}[\matrixfontsize][c]{,c15c,}
                    a  & b\\
                    c  & d
                    \end{IEEEeqnarraybox*}\right]%
=
                    \left[ \begin{IEEEeqnarraybox*}[\matrixfontsize][c]{,c15c,}
                    a  & K^{-2} b\IEEEstrut[10pt][2pt] \\
                    K^2 c  & d\IEEEstrut[10pt][1pt]
                    \end{IEEEeqnarraybox*}\right]\,.%
\end{equation}

\begin{defn}\label{defn:D_invariance}
A group $\mathscr{G<N}$ is \emph{$\mathscr{D}$-invariant} if all of the inner automorphisms \mbox{$\gam{\,\mathbf D}\in\gamma(\mathscr{D})$} fix the group $\mathscr{G}$; i.e., \mbox{$\gam{\,\mathbf D}\mathscr{G}=\mathscr{G}$,}  so that \mbox{$\gam{\,\mathbf D}|_{\mathscr{G}}\in\Aut(\mathscr{G})$}.  This is equivalent to saying that
$\mathscr{D}$ lies in the normalizer of $\mathscr{G}$ in $\mathscr{N}$:
\[ 
{\mathscr{D}} \;<\; N_{\mathscr{N}}(\mathscr{G}) \;\equiv\;
\left\{\mathbf{A}\in\mathscr{N} : \mathbf{A}\mathscr{G}\mathbf{A}^{-1} = \mathscr{G}\right\}\,.
\]
\end{defn}
\begin{lem}\label{lem:D-invariance}
Given an additive group $\mathscr{P}$ of lifting filters, the groups
$\upsilon(\mathscr{P})$ and $ \lambda(\mathscr{P})$
are $\mathscr{D}$-invariant if and only if $\mathscr{P}$ is closed under multiplication by $K^2$ for all $\mathbf{D}_K\in\mathscr{D}$.
\end{lem}
\pf Follows immediately from~(\ref{lifting_conjugation}).\hfill\qed

\subsection{Group Lifting Structures}\label{sec:Abelian:Structures}
The final component in a universe of lifting factorizations is  a set, $\mathfrak{B}$, of base filter banks from which other filter banks are lifted using lifting matrices in $\mathscr{U}$ and $\mathscr{L}$.  Using a set rather than a single base filter bank will allow us to prove uniqueness of base filter banks as well as uniqueness of lifting matrices. $\mathfrak{B}$ need not form a group, as will be seen in the case of HS filter banks.
We can now define a universe of lifting factorizations, called a \emph{group lifting structure.}

\begin{defn}\label{defn:LiftingStructures}
A \emph{group lifting structure}, $\mathfrak{S}$, is an ordered four-tuple, 
\[ \mathfrak{S}\equiv(\mathscr{D},\,\mathscr{U},\,\mathscr{L},\,\mathfrak{B})\,, \]
where $\mathscr{D}$ is a gain scaling group, $\mathscr{U}$ and $\mathscr{L}$ are  upper and lower triangular lifting matrix groups, and $\mathfrak{B}\subset\mathscr{N}$.  
The \emph{lifting cascade group},  $\mathscr{C}$, generated by $\mathfrak{S}$ is the subgroup of $\mathscr{N}$ generated by $\mathscr{U}$ and $\mathscr{L}$:
\begin{equation}\label{lifting_cascade_group}
\mathscr{C} \;\equiv\; \langle\mathscr{U\cup L}\rangle \;=\;
\left\{\rule[-1pt]{0pt}{12pt}\mathbf{S}_1\cdots\mathbf{S}_k \,:\, 
k\geq 1,\;\mathbf{S}_i\in\mathscr{U\cup L}\right\}\,. 
\end{equation}
We say  $\mathfrak{S}$ is a \emph{$\mathscr{D}$-invariant group lifting structure} if $\mathscr{U}$ and $\mathscr{L}$, and therefore $\mathscr{C}$, are $\mathscr{D}$-invariant groups.
\end{defn}

The set of all filter banks generated by $\mathfrak{S}$ is 
\[ 
\mathscr{DC}\mathfrak{B} \equiv 
\left\{\rule[-1pt]{0pt}{12pt}\mathbf{DCB} : 
\mathbf{D}\in\mathscr{D},\;\mathbf{C}\in\mathscr{C}\,,\;\mathbf{B}\in\mathfrak{B}\right\}\,.
\]
We will regard elements of $\mathscr{DC}\mathfrak{B}$ both as individual transfer matrices and as  partially factored lifting cascades of the form~(\ref{lift_from_B}) with gain matrix $\mathbf{D}_K\in\mathscr{D}$, lifting matrices $\mathbf{S}_i(z)\in\mathscr{U\cup L}$, and base $\mathbf{B}(z)\in\mathfrak{B}$.  The statement that $\mathbf{H}(z)$ ``has a (group) lifting factorization in $\mathfrak{S}$'' will mean that $\mathbf{H}\in\mathscr{DC}\mathfrak{B}$.  Moreover, since any group lifting factorization can be simplified via matrix multiplication into an irreducible lifting factorization with lifting matrices in the same  groups, we obtain the following elementary result.
\begin{prop}\label{prop:IrreducibleFactorizations}
$\mathbf{H}(z)$ has a group lifting factorization in $\mathfrak{S}$ if and only if it has an \emph{irreducible} lifting factorization in $\mathfrak{S}$.
\end{prop}

\begin{defn}\label{defn:ScaledLiftingGroup}
The \emph{scaled lifting group},  $\mathscr{S}$, generated by $\mathfrak{S}$ is the subgroup of $\mathscr{N}$ generated by $\mathscr{D}$ and $\mathscr{C}$:
\begin{IEEEeqnarray}{rCl}
\mathscr{S} & \equiv & \langle\mathscr{D\cup C}\rangle {}={}
\langle\mathscr{D\cup U\cup L}\rangle\IEEEnonumber \\%
&=&
\left\{\rule[-1pt]{0pt}{12pt}\mathbf{A}_1\cdots\mathbf{A}_k \,:\, 
k\geq 1,\;\mathbf{A}_i\in\mathscr{D\cup U\cup L}\right\}\,. \label{scaled_lifting_group}
\end{IEEEeqnarray}
\end{defn}
\rem
When two adjacent factors in~(\ref{scaled_lifting_group}) have the form $\mathbf{A}_i\mathbf{A}_{i+1}=\mathbf{SD}$ for $\mathbf{S}\in\mathscr{U\cup L}$ and $\mathbf{D}\in\mathscr{D}$, we can use~(\ref{intertwine_gain}) to rewrite the product as $\mathbf{SD}=\mathbf{DS}'$ where $\gam{\,\mathbf D}\mathbf{S}'=\mathbf{S}$.  In this way an arbitrary element of $\mathscr{S}$ can be written 
\begin{equation}\label{prod_dc}
\mathbf{A}_1\cdots\mathbf{A}_k = (\mathbf{D}_1\mathbf{D}_2\cdots)(\mathbf{S}'_1\mathbf{S}'_2\cdots)\,.
\end{equation}
If $\mathfrak{S}$ is $\mathscr{D}$-invariant then $\mathbf{S}'_i\in\mathscr{U\cup L}$ for all $i$, so~(\ref{prod_dc}) implies
\begin{equation}\label{S=DC}
\mathscr{S=DC}\quad\mbox{and}\quad\mathscr{DC}\mathfrak{B}=\mathscr{S}\mathfrak{B}
\quad\mbox{if $\mathfrak{S}$ is $\mathscr{D}$-invariant.}
\end{equation}
This shows that $\mathscr S$ embodies the group-theoretic aspects of $\mathscr{DC}\mathfrak{B}$ even if $\mathfrak B$ and $\mathscr{DC}\mathfrak{B}$ are not groups.  Work in preparation analyzes the group-theoretic structure of $\mathscr S$ and $\mathscr C$ in detail.

\section{Group Lifting Structures for Linear Phase Filter Banks}\label{sec:LinearPhase}
We now give examples of group lifting structures that parameterize the factorization of linear phase filter banks into linear phase lifting steps, based on the factorization theory in~\cite{BrisWohl06}.  It would also be nice to have examples of group lifting structures for paraunitary filter banks~\cite{Vaid93} since paraunitary transfer matrices naturally form a group, but lifting is not as well-suited for paraunitary matrices as it is for WS or HS matrices.  The problem is not existence of lifting factorizations: the theory in~\cite{DaubSwel98} ensures that one can always  factor paraunitary matrices into lifting cascades, and a three-step lifting structure is given in~\cite[Section~7.2]{DaubSwel98} for arbitrary $2\times 2$ rotations, yielding a general lifting parameterization for paraunitary filter banks via Givens rotations.  Rather, the problems stem from the fact that  individual lifting matrices are never paraunitary in their own right.  
This means that lifting factorization of paraunitary matrices necessarily takes place 
\emph{outside} of the paraunitary group.  Consequently, there does not appear to be any simple way of defining a group lifting structure that algebraically enforces the paraunitary property.  Moreover, it means that the paraunitary property is not robust with respect to quantization effects in finite-precision lifting implementations.  This is in marked contrast to the cases of WS and HS filter banks.

\subsection{Whole-Sample Symmetric Filter Banks}\label{sec:LinearPhase:WS}
In~\cite[Theorem~3]{BrisWohl06} it was shown that the family of delay-minimized WS filter banks with real impulse responses are defined by an intertwining relation, 
\begin{equation}\label{WS_intertwining}
\mathbf{H}(z^{-1}) = \bsy{\Lambda}(z)\mathbf{H}(z)\bsy{\Lambda}(z^{-1})\,,
\end{equation}
with $\bsy{\Lambda}(z)\equiv\diag(1,\,z^{-1})$.
Filter banks satisfying~(\ref{WS_intertwining}) form a subgroup of $\mathscr{F}$ called the \emph{WS group}.  
\begin{defn}\label{defn:uni_WS_group}
The \emph{unimodular WS  group}, $\mathscr{W}$, is the intersection of the WS group and  $\mathscr{N}$; i.e., $\mathscr{W}$ is the group of all real FIR transfer matrices that satisfy both~(\ref{DS_FIR_PR}) and~(\ref{WS_intertwining}).
\end{defn}
\begin{exmp}\label{exmp:uni_WS_lifting_structure}
\emph{Group lifting structure for the unimodular  WS group.}
Define the following groups of half-sample symmetric real Laurent polynomials:
\begin{IEEEeqnarray}{rCl}
\mathscr{P}_0 & \equiv& \left\{S(z)\in\mathbb{R}[z,z^{-1}] \,:\, S(z^{-1})=zS(z)\right\}\,,\label{WS_G0}\\
\mathscr{P}_1 & \equiv& \left\{S(z)\in\mathbb{R}[z,z^{-1}]  \,:\, S(z^{-1})=z^{-1}S(z)\right\}\,.\label{WS_G1}
\end{IEEEeqnarray}
According to~\cite[Theorem~9]{BrisWohl06}, every unimodular WS filter bank  factors completely  using lifting filters from $\mathscr{P}_0$ and $\mathscr{P}_1$.  Specifically, the lifting matrix groups are
\mbox{$\mathscr{U} \equiv\upsilon(\mathscr{P}_0)$} and $\mathscr{L} \equiv\lambda(\mathscr{P}_1)$.
Since  unimodular WS filter banks  factor completely (i.e., can be ``lifted from the lazy wavelet'') using matrices in $\mathscr{U}$ and $\mathscr{L}$, set $\mathfrak{B} \equiv\left\{\mathbf{I}\right\}$.
Let $\mathscr{D} \equiv\left\{\mathbf{D}_K : K\in\mathbb{R}^*\right\}$  
and $\mathfrak{S}_{\mathscr{W}}\equiv(\mathscr{D},\,\mathscr{U},\,\mathscr{L},\,\mathfrak{B})$, which is $\mathscr D$-invariant by Lemma~\ref{lem:D-invariance}.  
In terms of $\mathfrak{S}_{\mathscr{W}}$, 
the WS lifting factorization theorem~\cite[Theorem~9]{BrisWohl06} can be stated succinctly as
\begin{IEEEeqnarray*}{rCl}
\mathscr{W} &=& \mathscr{DC_W}\\
& = & \mathscr{S_W} \quad\mbox{by~(\ref{S=DC}),}
\end{IEEEeqnarray*}
where  $\mathscr{C_W\equiv\langle U\cup L\rangle}$ and  $\mathscr{S_W\equiv\langle D\cup C_W\rangle}$.  
\end{exmp}

\paragraph*{Irreversible {WS} Filter Banks in {JPEG~2000}}
JPEG~2000 Part~2~\cite[Annex~G]{ISO_15444_2} allows user-defined irreversible WS filter banks specified as irreducible lifting cascades in $\mathscr{S_W}$.  JPEG~2000 does not accommodate \emph{all} possible  filter banks in $\mathscr{S_W}$, however, because the standard imposes a  lowpass DC gain normalization requirement~\cite[Annex~G.2.1]{ISO_15444_2},~\cite{BrisWohl07}:
\begin{equation}\label{lowpass_DC_gain}
H_0(1)=1\,.
\end{equation}
The irreversible WS filter banks permitted by~\cite[Annex~G]{ISO_15444_2}  form a proper subset of $\mathscr{S_W}$: 
each unnormalized lifting cascade $\mathbf{E}(z)\in\mathscr{C_W}$ that satisfies $E_0(1)\neq 0$ must be paired with the unique gain scaling factor $K= E_0(1)$, which implies~(\ref{lowpass_DC_gain}).  See~\cite{BrisWohl07} for an analysis of JPEG~2000 gain normalization.

The WS filter banks satisfying the  normalization~(\ref{lowpass_DC_gain}) do \emph{not} form a subgroup of  $\mathscr{W}$.  For instance, the filter bank
\begin{equation}\label{WS_one_step} 
E_0(z)\equiv z+1+z^{-1}\quad\mbox{and}\quad E_1(z)\equiv z 
\end{equation}
has $\mathbf{E}(z)=\mathbf{S}(z)$ with lifting filter $S(z)=1+z^{-1}.$  The normalized polyphase matrix $\mathbf{H}(z)\equiv \mathbf{D}_3\mathbf{S}(z)$ satisfies~(\ref{lowpass_DC_gain}) but $\mathbf{H}^2(z)$ does not, hence such matrices do not form a group.

\begin{exmp}\label{exmp:rev_WS_lifting_structure}
\emph{Group lifting structure for the reversible  WS group.}
In contrast to the group $\mathscr{W}$, whose definition is independent of the notion of lifting, the group of \emph{reversible} WS filter banks, $\mathscr{W}_r$, is defined directly in terms of lifting~\cite{ZandiAllenEtal::Compression-with-reversible,CDSY98,TaubMarc02}.
Let $\mathbb{D}$ be the ring of dyadic  rationals,
\begin{equation}\label{dyadics}
\mathbb{D}\equiv \left\{\, k2^n : k,n\in\mathbb{Z}\,\right\}\,,
\end{equation}
and define  groups of dyadic symmetric Laurent polynomials:
\begin{IEEEeqnarray}{rCl}
\mathscr{P}_0 & \equiv& \left\{S(z)\in\mathbb{D}[z,z^{-1}] \,:\, S(z^{-1})=zS(z)\right\}\,,\label{WS_R_G0}\\
\mathscr{P}_1 & \equiv& \left\{S(z)\in\mathbb{D}[z,z^{-1}]  \,:\, S(z^{-1})=z^{-1}S(z)\right\}\,.\label{WS_R_G1}
\end{IEEEeqnarray}
The groups of reversible lifting matrices are
$\mathscr{U}_r \equiv\upsilon(\mathscr{P}_0)$ and $\mathscr{L}_r \equiv\lambda(\mathscr{P}_1)$.
Since gain scaling is not used in reversible filter bank implementations, let $\mathscr{D}_r \equiv\{\mathbf{I}\}$.  Set 
$\mathfrak{B}_r \equiv\left\{\mathbf{I}\right\}$
and $\mathfrak{S}_{\mathscr{W}_r}\equiv(\mathscr{D}_r,\,\mathscr{U}_r,\,\mathscr{L}_r,\,\mathfrak{B}_r)$ with $\mathscr{C}_{\mathscr{W}_r}\equiv\langle\mathscr{U}_r\cup \mathscr{L}_r\rangle$.
The \emph{reversible WS group} is the group \emph{defined} by
\[ \mathscr{W}_r \equiv \mathscr{C}_{\mathscr{W}_r} \,. \]
Note that $\mathscr{W}_r$, which is a subgroup of the unimodular WS group, $\mathscr{W}$, consists of \emph{linear} filter banks and does {not} include specifications for the nonlinear rounding operations involved in reversible \emph{implementations} of these filter banks.
\end{exmp}

\paragraph*{Reversible {WS} Filter Banks in {JPEG~2000}}  JPEG~2000 is more restrictive regarding  user-defined reversible WS filter banks in~\cite[Annex~G]{ISO_15444_2}  than it is regarding irreversible ones because reversible implementations may not use gain scaling operations.  This means that the normalization~(\ref{lowpass_DC_gain}) must be implicit in the choice of dyadic HS lifting filters.  For instance, the  5-tap/3-tap LeGall-Tabatabai spline wavelet analysis filter bank~\cite{LeGallTabataba::Subband-coding-digital}
is specified in~\cite[Annex~F.4.8.1]{ISO_15444_1} and~\cite[Annex~G.4.2.1]{ISO_15444_2} using the dyadic lifting factorization
\begin{equation}\label{LeGaTab}
\mathbf{H}(z) = 
        \left[ \begin{IEEEeqnarraybox*}[\matrixfontsize][c]{,c15c,}
                1  & (1+z^{-1})/4 \\
                0  & 1
        \end{IEEEeqnarraybox*}\right]
        \left[ \begin{IEEEeqnarraybox*}[\matrixfontsize][c]{,c15c,}
                1  & 0 \\
                -(z+1)/2  & 1
        \end{IEEEeqnarraybox*}\right]\,,
\end{equation}
which satisfies~(\ref{lowpass_DC_gain}).
Note that~(\ref{WS_one_step}) provides an example of a dyadic lifted WS filter bank that does \emph{not} satisfy~(\ref{lowpass_DC_gain}) without a gain scaling renormalization and therefore cannot be implemented reversibly within the constraints of~\cite{ISO_15444_2}.

\rem  While one can define reversible integer-to-integer filter banks by inserting rounding operations into floating-point lifting cascades~\cite{CDSY98}, this produces  finite-precision sensitivities that could cause problems in communications applications where data is encoded on one platform and decoded on another.  For this reason, prevailing practice (e.g.,~\cite{ISO_15444_2}) has been to restrict reversible implementations to dyadic filter banks.  (A side benefit  is the opportunity to replace floating point arithmetic with integer arithmetic and bit-shifts.)  This creates difficulties in factorization theory, however, since $\mathbb{D}[z,z^{-1}]$ is not a Euclidean domain. This means that lifting factorization of a transfer matrix over $\mathbb{D}[z,z^{-1}]$ may not yield lifting filters in $\mathbb{D}[z,z^{-1}]$.  For instance, if $b=2$ and $c=3$ then the transfer matrix $\mathbf{H}(z)$ in Example~\ref{exmp:nonunique_3_factors} of Section~\ref{sec:Filter:Irreducible} has matrix components in $\mathbb{D}[z,z^{-1}]$, but  the  factorization 
\[ \mathbf{H}(z)=\mathbf{S}'_2(z)\,\mathbf{S}'_1(z)\,\mathbf{S}'_0(z) \]
does \emph{not} have dyadic lifting filters even though it could have been constructed using the Euclidean algorithm.

\subsection{Half-Sample Symmetric Filter Banks}\label{sec:LinearPhase:HS}
Concentric delay-minimized HS filter banks (in which both impulse responses are centered at $-1/2$) are characterized in~\cite[eqn.~(37)]{BrisWohl06}  by the relation
\begin{equation}\label{HS_symmetry}
\mathbf{H}(z^{-1}) = \mathbf{L}\,\mathbf{H}(z)\,\mathbf{J}\,,
\end{equation}
where
\[
\mathbf{L} \equiv 
        \left[ \begin{IEEEeqnarraybox*}[\matrixfontsize][c]{,c15c,}
	1  &  0 \\
	0  & -1  
	\end{IEEEeqnarraybox*}\right]
\quad\mbox{and}\quad
\mathbf{J} \equiv 
        \left[ \begin{IEEEeqnarraybox*}[\matrixfontsize][c]{,c15c,}
	0  &  1 \\
	1  & 0  
	\end{IEEEeqnarraybox*}\right] \,.
\]
Unlike the case of WS filter banks,~(\ref{HS_symmetry}) does \emph{not} define a matrix group, which is reflected in the fact that the base filter banks for the HS group lifting structure do not form a group.
\begin{defn}\label{defn:conc_HS_class}
The \emph{unimodular HS class}, $\mathfrak{H}$, is the class of all real FIR transfer matrices that satisfy both~(\ref{DS_FIR_PR}) and~(\ref{HS_symmetry}).
\end{defn}
\begin{exmp}\label{exmp:uni_HS_lifting_structure}
\emph{Group lifting structure for the unimodular HS class.}
Let  $\mathscr{D}\equiv\left\{\mathbf{D}_K : K\in\mathbb{R}^*\right\}$.
Define the following group of whole-sample antisymmetric  real Laurent polynomials:
\begin{equation}\label{WA_polys}
\mathscr{P} \equiv \left\{S(z)\in\mathbb{R}[z,z^{-1}] \,:\, S(z^{-1})=-S(z)\right\}\,.
\end{equation}
Let $\mathscr{U}\equiv\upsilon(\mathscr{P})$ and $\mathscr{L}\equiv\lambda(\mathscr{P})$.
The set of concentric equal-length base filter banks for the unimodular HS class is
\begin{equation}\label{B_HS}
\mathfrak{B_H}\equiv \left\{\mathbf{B}\in\mathfrak{H}:\order(B_0)=\order(B_1)\right\}. 
\end{equation}
Set $\mathfrak{S_H}\equiv(\mathscr{D},\,\mathscr{U},\,\mathscr{L},\,\mathfrak{B_H})$ and  
$\mathscr{S}_{\mathfrak{H}}\equiv\mathscr{\langle D\cup U\cup L\rangle}$.
Then  $\mathfrak{S_H}$ is $\mathscr D$-invariant  by Lemma~\ref{lem:D-invariance}, and the HS  lifting factorization theorem~\cite[Theorem~14]{BrisWohl06} can be stated succinctly as
\[ \mathfrak{H} = \mathscr{S}_{\mathfrak{H}}\mathfrak{B_H} \,. \]
\end{exmp}
\begin{exmp}\label{exmp:rev_HS_lifting_structure}
\emph{Group lifting structure for the reversible  HS class.}
As with WS filter banks, the class  of \emph{reversible} HS filter banks, $\mathfrak{H}_r$, is defined in terms of lifting.  
Define the following group of dyadic antisymmetric Laurent polynomials:
\begin{equation}\label{rev_WA_polys}
\mathscr{P} \equiv \left\{S(z)\in\mathbb{D}[z,z^{-1}] \,:\, S(z^{-1})=-S(z)\right\}\,.
\end{equation}
The groups of lifting matrices are
$\mathscr{U}_r\equiv\upsilon(\mathscr{P})$ and $\mathscr{L}_r\equiv\lambda(\mathscr{P})$.

Defining the set  $\mathfrak{B}_{\mathfrak{H}_r}$ of reversible HS base filter banks  is problematic.  As in~(\ref{B_HS}) they  must be concentric equal-length HS filter banks, but  in light of the remarks at the end of Example~\ref{exmp:rev_WS_lifting_structure} we cannot simply take the set of all such filter banks with dyadic coefficients because we have no guarantee that they will have lifting factorizations over $\mathbb{D}[z,z^{-1}]$.  Instead, define  $\mathfrak{B}_{\mathfrak{H}_r}$ as the set of all matrices in~$\mathfrak{B}_{\mathfrak{H}}$ that \emph{have}  lifting factorizations over  $\mathbb{D}[z,z^{-1}]$.  
(Any  scaling matrix, $\mathbf{D}_K$, for which $K=2^n$ can be factored into dyadic lifting steps using the formulas in~\cite[Section~7.3]{DaubSwel98}.
Also, note that by~\cite[Theorem~13]{BrisWohl06} any lifting factorization of a concentric equal-length, delay-minimized HS base filter bank requires some lifting filters that are \emph{not} whole-sample antisymmetric.)

Let $\mathscr{D}_r\equiv\{\mathbf{I}\}$
and set $\mathfrak{S}_{\mathfrak{H}_r}\equiv(\mathscr{D}_r,\,\mathscr{U}_r,\,\mathscr{L}_r,\,\mathfrak{B}_{\mathfrak{H}_r})$.
The \emph{reversible HS class} is  defined to be
\[ \mathfrak{H}_r \equiv \mathscr{C}_{\mathfrak{H}_r}\mathfrak{B}_{\mathfrak{H}_r} \,. \]
\end{exmp}

\begin{exmp}\label{exmp:ELASF}
\emph{The ELASF family of M.\ Adams.}  The Ph.D.\ dissertation of M.\ Adams~\cite{Adams02,AdamsWard:03:Symmetric-extension-compatible-reversible-integer-to-integer}
defines a class of reversible  lifted HS filter banks referred to as the \emph{even-length analysis/synthesis filter} (ELASF) family.  Adams studied constraints on the rounding rules used in reversible  filter bank implementations  resulting from the requirement that a reversible implementation yield symmetric or antisymmetric integer subbands in  symmetric pre-extension schemes~\cite{WohlBris08}.  Ignoring issues related to  rounding rules, which are not relevant to this paper, we can construct a group lifting structure for the linear (non-rounded) ELASF family.
All ELASF filter banks are  lifted from the Haar~(\ref{Haar}) so $\mathfrak{B}\equiv\{\mathbf{H}_{haar}\},$ and all are reversible so  $\mathscr{D}_r\equiv\{\mathbf{I}\}$.
The lifting filters for the ELASF family are the same antisymmetric dyadic filters~(\ref{rev_WA_polys}) used in   $\mathfrak{S}_{\mathfrak{H}_r}$.
Let $\mathfrak{S}_{\ssst ELASF}$ be the corresponding group lifting structure.

Adams also specifies the reversible lifting factorization~(\ref{Haar_lifting_steps}) for the Haar base filter bank, and when the first WA lifting step ($\mathbf{S}_0(z)$ in~(\ref{lift_from_B})) is upper triangular he advocates combining it with the upper triangular step in~(\ref{Haar_lifting_steps}) before the first rounding operation to minimize quantization effects (see~\cite[Figure~3]{AdamsWard:03:Symmetric-extension-compatible-reversible-integer-to-integer}).  
Since we are only interested in linear (non-rounded) filter banks, however, we regard (linear) ELASF filter banks as being defined by a partially factored lifting cascade~(\ref{lift_from_B}) lifted from a single base filter bank, $\mathbf{H}_{haar}$, using antisymmetric dyadic lifting filters.  The set of all (linear) ELASF filter banks, $\mathscr{C}\mathfrak{B}$, generated by  $\mathfrak{S}_{\ssst ELASF}$ is therefore a subset of the reversible HS class, $\mathfrak{H}_r$, and every group lifting factorization in $\mathfrak{S}_{\ssst ELASF}$ is also a group lifting factorization in $\mathfrak{S}_{\mathfrak{H}_r}$.
\end{exmp}

\section{A Uniqueness Theory for Lifting Cascades}\label{sec:Uniqueness}
We now establish sufficient conditions on a  group lifting structure to imply that all of the matrix factors (the lifting matrices, $\mathbf{S}_i(z)$, and the gain scaling and base matrices) in irreducible group lifting factorizations are unique modulo one trivial degree of freedom.

\subsection{The Order-Increasing Property}
\label{sec:Uniqueness:Order-Increasing}
From an engineering perspective, factorization~(\ref{identity_lift}) is of no practical interest because it  doesn't synthesize a high-order filter bank from low-order building blocks.  (For instance, (\ref{identity_lift}) is not a factorization one would arrive at via the Euclidean algorithm.)  
Writing  nonunique lifting factorizations  as factorizations of the identity, as in~(\ref{identity_lift}) and~(\ref{identity_lift2}), raises an interesting point.
Given a lifting of the identity, if some partial product $\mathbf{E}^{(n)}(z)$  has nonzero order  then the order of subsequent partial products must eventually \emph{decrease} so that the final product  (which is $\mathbf{I}$) has order zero.
This suggests that lifting structures that only generate ``order-increasing'' cascades will also generate \emph{unique} factorizations, an idea  that will be made rigorous in Theorem~\ref{thm:unique_factorization}.  
Evidently, though, the notion of order reduction embodied in the Euclidean algorithm \emph{is not sufficiently strong} to imply uniqueness  when  factorizations are constructed via the Euclidean algorithm as in~\cite{DaubSwel98}.
We therefore need to define an appropriate notion of ``order-increasing'' lifting cascades.  The key is to focus on the \emph{polyphase} orders of the intermediate transfer matrices $\mathbf{E}^{(n)}(z)$ rather than on the polynomial orders of the intermediate scalar filters.

\begin{defn}\label{defn:OrderIncreasing}
A  lifting cascade~(\ref{lift_from_B}) is {\em strictly polyphase order-increasing\/} (usually shortened to just \emph{order-increasing}) if the order~(\ref{FB_order}) of each intermediate polyphase matrix~(\ref{recursive_cascade}) is strictly greater than the order of its predecessor:
\[ 
\order\left(\mathbf{E}^{(n)}\right) > \order\left(\mathbf{E}^{(n-1)}\right)\quad\mbox{for $0\leq n < N$.}
\]
A group lifting structure, $\mathfrak{S}$, will be  called order-increasing if every irreducible cascade in 
\mbox{$\mathscr{C}\mathfrak{B}$} is order-increasing.
\end{defn}
\begin{exmp}\label{exmp:Non-Order-Increasing}
\emph{WS filter bank with a \textbf{non}-order-increasing HS lifting cascade.}
Define the \emph{causal} (nonunimodular) lazy wavelet filter bank using $B_0(z)\equiv 1$ and $B_1(z)\equiv z^{-1}$;
i.e., instead of lifting from the identity as in Example~\ref{exmp:uni_WS_lifting_structure}  we are lifting from
\begin{equation}\label{causal_lazy}
\mathbf{B}(z) = 
        \left[ \begin{IEEEeqnarraybox*}[\matrixfontsize][c]{,c15c,}
                1   & 0\IEEEstrut[10pt][1pt] \\
                0   & z^{-1}\IEEEstrut[10pt][1pt]
    \end{IEEEeqnarraybox*}\right]
    \,,
\end{equation}
whose order is one.  Lift ${B_1}(z)$ to a higher-order WS filter, $E^{(0)}_1(z)$, using the HS lifting filter $S_0(z)\equiv 1+z^{-1}$:
\[ E^{(0)}_1(z)\;=\;B_1(z)+S_0(z^2)B_0(z)\;=\;1+z^{-1}+z^{-2}\,. \]
The symmetry of $S_0$ is chosen to leave $e^{(0)}_1$  symmetric about $n=1$, just like $b_1$.
The lifted polyphase matrix~(\ref{recursive_cascade}) is
\begin{IEEEeqnarray*}{rCl}
\mathbf{E}^{(0)}(z)
&=&
	\left[ \begin{IEEEeqnarraybox*}[\matrixfontsize][c]{,c15c,}
                1   & 0 \IEEEstrut[10pt][1pt]\\
                1+z^{-1}   & 1\IEEEstrut[10pt][1pt]
	\end{IEEEeqnarraybox*}\right]%
	\left[ \begin{IEEEeqnarraybox*}[\matrixfontsize][c]{,c15c,}
                1   & 0\IEEEstrut[10pt][1pt] \\
                0   & z^{-1}\IEEEstrut[10pt][1pt]
	\end{IEEEeqnarraybox*}\right]    \\%
&=&
	\left[ \begin{IEEEeqnarraybox*}[\matrixfontsize][c]{,c15c,}
                1   & 0\IEEEstrut[10pt][1pt]  \\
                1+z^{-1}   & z^{-1}\IEEEstrut[10pt][1pt] 
	\end{IEEEeqnarraybox*}\right]\,,
\end{IEEEeqnarray*}
whose polyphase order is still one (no increase) even though the order of the highpass scalar filter increased from zero to two.  Despite the similarities between this example and  Example~\ref{exmp:uni_WS_lifting_structure},
we show in~\cite{Bris09b}  that the group lifting structure  for the unimodular WS group, $\mathscr{W}$, defined in Example~\ref{exmp:uni_WS_lifting_structure}   \emph{is} strictly polyphase order-increasing, a somewhat nontrivial task.
\end{exmp}

\subsection{A General Uniqueness Theorem}
\label{sec:Uniqueness:General}
As noted above, any  lifting cascade  can be  simplified to irreducible form by matrix multiplication.  Since $\mathscr{U}$ and $\mathscr{L}$ are groups,  every transfer matrix generated by $\mathfrak{S}$ has an irreducible lifting factorization in \mbox{$\mathscr{D}\mathscr{C}\mathfrak{B}$.}  Our main result shows that such factorizations are ``unique modulo rescaling'' if $\mathfrak{S}$  is order-increasing (per Definition~\ref{defn:OrderIncreasing}) and $\mathscr{D}$-invariant.
\begin{thm}[Uniqueness of  lifting factors]
\label{thm:unique_factorization}
Suppose that 
$ \mathfrak{S} = (\mathscr{D},\,\mathscr{U},\,\mathscr{L},\,\mathfrak{B}) $
is a $\mathscr D$-invariant, order-increasing group lifting structure.  Let $\mathbf{H}(z)$ be a transfer matrix generated by $\mathfrak{S}$, and suppose we are given two irreducible group lifting factorizations of $\mathbf{H}(z)$ in \mbox{$\mathscr{D}\mathscr{C}\mathfrak{B}$}:
\begin{eqnarray}
\mathbf{H}(z) & = & \mathbf{D}_K\,\mathbf{S}_{N-1}(z) \cdots\mathbf{S}_0(z)\, \mathbf{B}(z)  
                \label{unprimed_cascade} \\
        & = & \mathbf{D}_{K'}\,\mathbf{S}'_{N'-1}(z)\cdots \mathbf{S}'_0(z)\,\mathbf{B}'(z)\;.
                \label{primed_cascade}
\end{eqnarray}
Then~(\ref{unprimed_cascade}) and~(\ref{primed_cascade}) satisfy the following three properties:
\begin{IEEEeqnarray}{rCl}
N' & = & N\,, \label{NprimeEqualsN}\\
\mathbf{B}'(z) & = & \mathbf{D}_{\alpha}\,\mathbf{B}(z)\quad\mbox{where\ }\alpha\equiv K/K', \label{defn_alpha}\\
\mathbf{S}'_i(z) & = & \gamma_{\!\alpha}\mathbf{S}_i(z)\quad\mbox{for $i=0,\ldots,N-1$.} \label{almost_unique_factors}
\end{IEEEeqnarray} 

If, in addition, $\mathbf{B}(z)$ and $\mathbf{B}'(z)$ share a nonzero matrix entry at some point $z_0$ then the factorizations~(\ref{unprimed_cascade}) and~(\ref{primed_cascade}) are \emph{identical;} i.e., $K'=K$, $\mathbf{B}'(z)=\mathbf{B}(z)$, and
\begin{equation}\label{unique_factors}
\mathbf{S}'_i(z) = \mathbf{S}_i(z)\quad \mbox{for $i=0,\ldots,N-1$.}
\end{equation}
It also follows that $K'=K$ if either of the scalar filters, $B_0(z)$ or $B_1(z)$, shares a nonzero value with its primed counterpart; e.g.,  if the base filter banks have equal lowpass DC responses.
\end{thm}
\rem
The type of lifting equivalence described above  is the subject of the following definition.
\begin{defn}\label{defn:ModuloRescaling}
Two factorizations of $\mathbf{H}(z)$ that satisfy (\ref{NprimeEqualsN})--(\ref{almost_unique_factors}) are said to be \emph{equivalent modulo rescaling.}  If \emph{all} irreducible group lifting factorizations of $\mathbf{H}(z)$ are equivalent modulo rescaling for \emph{every} $\mathbf{H}(z)$ generated by $\mathfrak{S}$, we say that irreducible factorizations in $\mathfrak{S}$ are \emph{unique modulo rescaling.} 
\end{defn}

We now prove Theorem~\ref{thm:unique_factorization}.
\pf
Without loss of generality, assume that $\order(\mathbf{B}')\geq\order(\mathbf{B}).$  Equate~(\ref{unprimed_cascade}) and~(\ref{primed_cascade}), and combine the gain scaling matrices on the unprimed side:
\begin{IEEEeqnarray}{l}
\mathbf{D}_{\alpha}\,\mathbf{S}_{N-1}(z) \cdots\mathbf{S}_0(z)\,\mathbf{B}(z) =  \IEEEnonumber\\
\qquad\qquad\qquad\mathbf{S}'_{N'-1}(z)\cdots \mathbf{S}'_0(z)\,\mathbf{B}'(z)\,,\label{simplify_1}
\end{IEEEeqnarray}
where $\alpha= K/K'$ and $\mathbf{D}_{\alpha}\in\mathscr{D}$.
Move the lifting matrices $\mathbf{S}_i(z)$ from the left-hand side of~(\ref{simplify_1}) to the right-hand side using~(\ref{intertwine_gain}) and~(\ref{gamma_Ainv}):
\begin{IEEEeqnarray}{l}
\mathbf{D}_{\alpha}\,\mathbf{B}(z) = \IEEEnonumber\\
\quad\gamma_{\!\alpha}\mathbf{S}^{-1}_0(z)\cdots \gamma_{\!\alpha}\mathbf{S}^{-1}_{N-1}(z)\,
\mathbf{S}'_{N'-1}(z)\cdots \mathbf{S}'_0(z)\,\mathbf{B}'(z)\,.\quad\label{simplify_2}
\end{IEEEeqnarray}
Since $\mathscr{U}$ and $\mathscr{L}$ are $\mathscr{D}$-invariant groups, the matrices $\gamma_{\!\alpha}\mathbf{S}^{-1}_i(z)$ are still in $\mathscr{U}\cup\mathscr{L}$.  

Simplify~(\ref{simplify_2}) to irreducible form via matrix multiplication.  Since~(\ref{unprimed_cascade}) and~(\ref{primed_cascade}) are irreducible,  simplification can only occur if $\gamma_{\!\alpha}\mathbf{S}^{-1}_{N-1}(z)$ and $\mathbf{S}'_{N'-1}(z)$ have the same update characteristic.  If so, combine them into a single lifting matrix, and if it reduces to the identity, i.e., if 
\[ \gamma_{\!\alpha}\mathbf{S}_{N-1}(z)=\mathbf{S}'_{N'-1}(z)\,, \]
 then combine $\gamma_{\!\alpha}\mathbf{S}^{-1}_{N-2}(z)$ and $\mathbf{S}'_{N'-2}(z)$, etc.  Simplification will eventually terminate in an irreducible cascade,
\begin{equation}\label{simplify_3}
\mathbf{D}_{\alpha}\,\mathbf{B}(z) = 
\mathbf{S}''_{M-1}(z)\cdots \mathbf{S}''_0(z)\,\mathbf{B}'(z)\,,
\end{equation}
for some $M\leq N+N'$.  The cascade on the right-hand side of~(\ref{simplify_3}) is still in \mbox{$\mathscr{C}\mathfrak{B}$} since $\mathbf{S}''_i(z)\in\mathscr{U\cup L}$ for all $i$.  If $M>0$ then the order-increasing property of $\mathfrak{S}$ implies that \mbox{$\order(\mathbf{B})>\order(\mathbf{B}')$,} a contradiction.  Therefore we must have $M=0$, which means we have  pairwise cancellation of all of the matrices $\gamma_{\!\alpha}\mathbf{S}^{-1}_i(z)$ and $\mathbf{S}'_i(z)$ in the reduction of~(\ref{simplify_2}) to irreducible form.
This  proves uniqueness modulo rescaling; i.e., conclusions~(\ref{NprimeEqualsN}), (\ref{defn_alpha}) and~(\ref{almost_unique_factors}).

If  $B_{ij}(z_0)=B'_{ij}(z_0)\neq 0$, then~(\ref{defn_alpha}) implies $K'=K$ and~(\ref{unique_factors}) follows.  
The case of $B_i(z)$ sharing a value with its primed counterpart, e.g.,   $B_0(1)=B'_0(1)\neq 0$, is similar.
\hfill\qed
%

\section{Conclusions}\label{sec:Concl}
A new algebraic framework, the group lifting structure, has been introduced for studying lifting factorizations of FIR perfect reconstruction filter banks.  The primary focus of the paper is the uniqueness of the matrix factors arising in lifting factorizations.
With no constraints on the lifting process, Proposition~\ref{prop:Nonuniqueness} shows that any FIR perfect reconstruction filter bank can be irreducibly lifted from any other FIR perfect reconstruction filter bank  in infinitely many different ways.  
Virtually all such factorizations are mathematical pathologies, however, and are of no interest for engineering applications.

To impose  constraints on the universe of feasible lifting factorizations for a given class of filter banks, we define  group lifting structures in terms of abelian groups of lower and upper triangular lifting matrices, an abelian group of gain scaling matrices, and  sets (not necessarily  groups) of base filter banks.  
If a group lifting structure satisfies a polyphase order-increasing hypothesis and if the groups of  lifting matrices are invariant under the action of the gain scaling group, the main result of the paper, Theorem~\ref{thm:unique_factorization}, shows that irreducible group lifting factorizations  are unique modulo, at most, rescaling. 

Theorem~\ref{thm:unique_factorization} is used in~\cite{Bris09b} to prove uniqueness results for  both reversible and irreversible lifting factorizations of WS and HS filter banks parameterized by the linear phase group lifting structures defined in Section~\ref{sec:LinearPhase} of the present paper.  This scope includes  the specification of  WS filter banks in JPEG~2000  and M.\ Adams' ELASF class of reversible HS filter banks.  The key to these proofs is establishing the order-increasing property for the associated group lifting structures.


%
\begin{IEEEbiography}[{\includegraphics[width=1in,height=1.25in,clip,keepaspectratio]{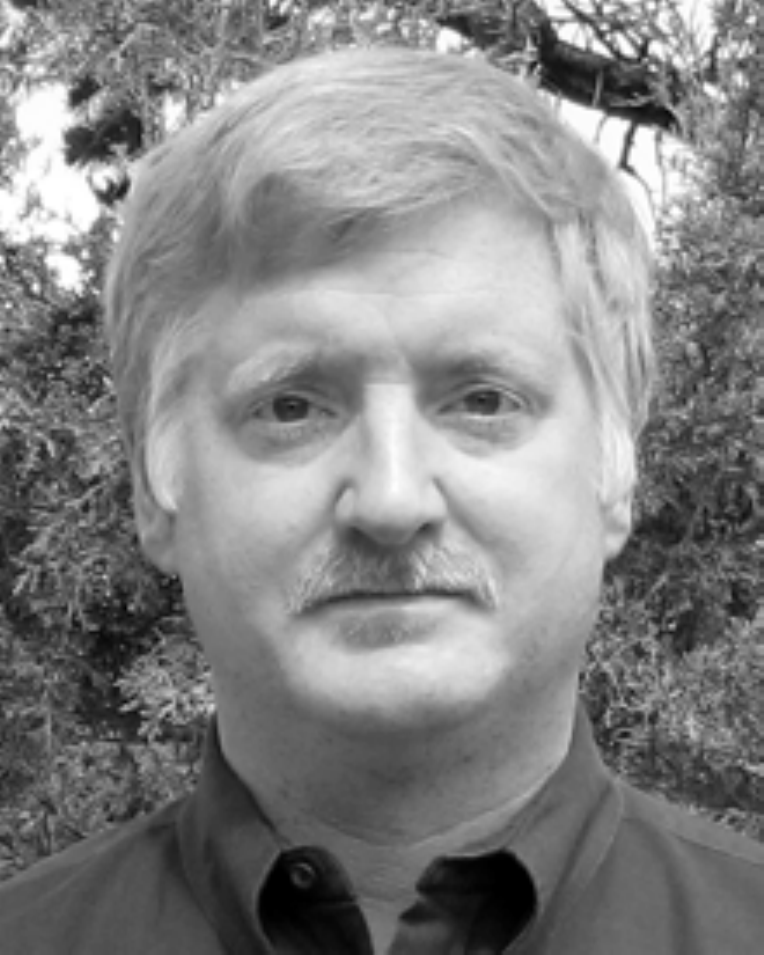}}]{Christopher M.\ Brislawn} (M'91--SM'05) received the B.S.\ degree in 1982 from Harvey Mudd College, Claremont, CA, and the Ph.D.\ degree in 1988 from the University of Colorado---Boulder, both in mathematics.

He was a Visiting Assistant Professor  at the University of Southern California from 1989 to 1990.  He joined Los Alamos National Laboratory (LANL), Los Alamos, NM,  as a postdoc in 1990 and became a permanent staff member in 1993.  Currently he is a  Scientist in Group CCS-3 of the Computer, Computational and Statistical Sciences Division.  His research interests include wavelet transforms, digital filter banks, communications coding, and statistical signal processing.  From 1991 to 1993 he coauthored the Wavelet/Scalar Quantization Specification for compression of digitized fingerprint images with the U.S.\ Federal Bureau of Investigation.  From 1999 to 2003 he served as LANL's Principal Member in Working Group L3.2 of the International Committee for Information Technology Standards   and led a LANL team that participated in writing the ISO/IEC JPEG~2000 standard (ISO 15444-x).  His team worked on Parts~1 and~2 of the standard, and he coauthored the proposal to create JPEG~2000 Part~10 (Extensions for Three-Dimensional Data), serving as the first editor of Part~10.  In 2007--2008 he represented LANL on  the Motion Imagery Standards Board for the National Geospatial Intelligence Agency.  In addition to his technical work, Dr.\ Brislawn has mentored 15 graduate students and 4 postdocs at LANL and has co-supervised one Ph.D.\ dissertation for the University of Texas---Austin.

Dr.\ Brislawn is also a member of the American Mathematical Society.
\end{IEEEbiography}
\end{document}